\documentclass[journal=jacsat,manuscript=article]{achemso}

\usepackage{chemformula}
\usepackage{upgreek}

\author{Sylvia Koerfer}
\author{Bianca Di{\ss}mann}
\author{Norman Schier}
\affiliation[RWTH]
{Institute of Physical Chemistry, RWTH Aachen University, Landoltweg 2, Aachen 52056, Germany}
\author{Han-Ill Yoo}
\affiliation[SNU]
{Department of Materials Science and Engineering, Seoul National University, Seoul 151-744, Republic of Korea}
\author{Manfred Martin}
\affiliation[RWTH]
{Institute of Physical Chemistry, RWTH Aachen University, Landoltweg 2, Aachen 52056, Germany}
\author{Roger A. De Souza}
\email{desouza@pc.rwth-aachen.de}
\phone{+49 (0)241 8094729}
\fax{+49 (0)241 8092128}
\affiliation[RWTH]
{Institute of Physical Chemistry, RWTH Aachen University, Landoltweg 2, Aachen 52056, Germany}

\title[Ba Diffusion in Perovskite BaTiO$_3$]
  {The Diffusion Kinetics of Ba Cations in Perovskite BaTiO$_3$: A Combined Tracer Diffusion and Metadynamics Study}


\begin{document}







\begin{abstract}
  Tracer diffusion experiments and metadynamics (MtD) simulations were used to study the diffusion of Ba cations in the cubic phase of the perovskite oxide BaTiO$_3$. $^{130}$BaTiO$_3$ thin films were used as diffusion sources to introduce barium tracer diffusion profiles into single-crystal samples at temperatures $1348 \leq T/\mathrm{K} \leq 1498$. The $^{130}$Ba profiles were determined by time-of-flight secondary ion mass spectrometry, and then analysed to yield Ba tracer diffusion coefficients ($D_\mathrm{Ba}^\ast$). MtD simulations were performed in order to obtain barium-vacancy diffusion coefficients ($D_\mathrm{v_{Ba}}$) for selected vacancy mechanisms as a function of temperature. $D_\mathrm{v_{Ba}}$ is predicted to be increased significantly by an adjacent oxygen vacancy, and even more, by an adjacent titanium vacancy. From the combined consideration of $D_\mathrm{Ba}^\ast$ and $D_\mathrm{v_{Ba}}$, we conclude that Ba diffusion in these samples occurred most probably by the migration of defect associates, and not by the migration of isolated barium vacancies. More generally, our results draw attention to the dangers of relying solely on activation enthalpies to interpret diffusion data.
\end{abstract}

\section{Introduction}\label{sec_intro}

\ch{BaTiO3} is arguably the preeminent electroceramic perovskite, if not because of its rich variety of interesting physical and chemical properties, then because of its application as a dielectric medium in multi-layer ceramic capacitors (MLCC), of which more than $10^{12}$ devices are produced annually worldwide. Each new smartphone contains more than $10^{3}$ MLCCs; each new electric vehicle, more than $10^{4}$ of them \cite{Hong.2019}. 

Much research on \ch{BaTiO3} has been devoted to investigating the behaviour of its point defects \cite{Eror.1978, Chan.1981, Nowotny.1991, Tsur.2001, Yoo.2002, Smyth.2003, Erhart.2007, Erhart.2008, Choi.2012, Noguchi.2021, Randall.2022}, with a view to understanding its physical and chemical properties and also to improving its performance in MLCC devices. Of all the important point defects, oxygen vacancies have attracted the most attention: they are generally present in the highest concentrations and they exhibit significant mobility, even at room temperature. It is this combination that adversely affects the performance of MLCC devices \cite{Waser.1990a, Waser.1990b, Baiatu.1990, Rodewald.2000, Wang.2016, Yousefian.2025}. Cation vacancies, in contrast, have received relatively little attention, partly because their concentrations are generally orders of magnitude lower, and partly because they are immobile at room temperature, showing appreciable mobilities only at very high temperatures (above, say, $1000\;^\circ$C). Nevertheless, cation vacancies are important defects because they allow cation diffusion to occur, and cation diffusion is required, to synthesise \ch{BaTiO3} from its binary oxides, to sinter a \ch{BaTiO3} ceramic body or device from a powder, and to condition a \ch{BaTiO3} powder or ceramic before use to achieve optimal performance. For the sake of completeness we note that there are a host of degradation processes, such as creep, grain growth, kinetic unmixing, and the segregation and accumulation of cations at extended defects, that depend on cation diffusion. Although these processes are of no relevance for \ch{BaTiO3} during device operation at and around room temperature, they are important during fabrication and they are important for related perovskite materials (e.g.~those based on \ch{BaZrO3} or \ch{SrTiO3}) used in elevated-temperature electrochemical devices.

\begin{figure}[h]
\centering
\includegraphics{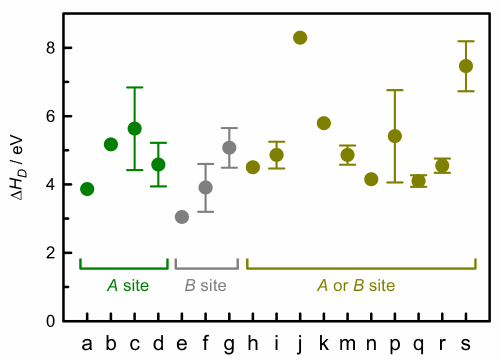}
\caption{Experimentally determined activation enthalpies of cation self-diffusion in perovskite \ch{BaTiO3}. \textit{A} site refers to studies in which Ba or Sr diffusion was probed; \textit{B} site, to studies in which Ti or Zr diffusion was probed; and \textit{A} or \textit{B} site, to data extracted from sintering, grain-growth or creep studies: a \cite{Garcia.1953}, b \cite{Kitahara.1999}, c \cite{Koerfer.2008}, d \cite{Sazinas.2017}, e \cite{Garcia.1953}, f \cite{Preis.2006}, g \cite{Koerfer.2008}, h \cite{Nomura.1956}, i \cite{Anderson.1965}, j \cite{Carry.1986}, k \cite{Genuist.1988}, m \cite{Beauchesne.1989}, n \cite{Xue.1990}, p \cite{Lin.2000}, q \cite{Lin.2000}, r \cite{Kambale.2014}, s \cite{Park.2015}.}
\label{fig_summaryliterature}
\end{figure}

Regardless of whether one considers fabrication or degradation processes, a detailed fundamental understanding of cation diffusion in \ch{BaTiO3} is lacking. Fig.~\ref{fig_summaryliterature} is a summary of activation enthalpies of cation diffusion, $\Delta H_D$, determined experimentally for \ch{BaTiO3}, either directly from tracer or impurity diffusion experiments for \textit{A}-site species \cite{Garcia.1953, Kitahara.1999, Koerfer.2008, Sazinas.2017} or for \textit{B}-site species \cite{Garcia.1953, Preis.2006, Koerfer.2008}, or indirectly from some process (such as interdiffusion \cite{Nomura.1956}, sintering \cite{Anderson.1965, Genuist.1988, Xue.1990, Lin.2000}, creep \cite{Carry.1986, Beauchesne.1989, Park.2015} or grain growth \cite{Kambale.2014}) that depends on cation diffusion, and thus could refer to either Ba or Ti diffusion (and may refer to grain-boundary, rather than bulk, diffusion). $\Delta H_D$ is seen to vary between 3~eV and 8~eV, with almost all values falling in a band between 4~eV and 6~eV, regardless of whether \textit{A}-site or \textit{B}-site diffusion is considered. Indeed, it is striking that there is no clear separation of the data into two groups, one corresponding to the diffusion of large, divalent cations on the \textit{A}-site sublattice, and the other, to the diffusion of small, tetravalent cations on the \textit{B}-site sublattice. This behaviour ($\Delta H_{D_A} \approx \Delta H_{D_B}$) is not specific to \ch{BaTiO3} but has also been observed for other \ch{$AB$O3} perovskites, such as \ch{La_{0.9}Sr_{0.1}Ga_{0.9}Mg_{0.1}O_{2.9}} \cite{Schulz.2003}, \ch{La_{0.9}Sr_{0.1}FeO_{3-$\delta$}} \cite{Waernhus.2007}, \ch{Ba_{0.5}Sr_{0.5}Co_{0.8}Fe_{0.2}O_{3-$\delta$}} \cite{Harvey.2012}, \ch{La_{0.6}Sr_{0.4}CoO_{3-$\delta$}} \cite{Kubicek.2014} and \ch{SrTiO3} \cite{Gries.2020}.

Part of the problem of understanding cation diffusion in perovskites is that even a tracer diffusion experiment, which probes the diffusion of a specific cation at chemical equilibrium, does not on its own yield sufficient information for a detailed analysis. A measured tracer diffusion coefficient of a cation in a perovskite is, namely, the product of the diffusivity of cation vacancies $D_\mathrm{v}$, the site fraction of cation vacancies $n_\mathrm{v}$ and the tracer correlation factor $f^\ast_\mathrm{v}$:
\begin{equation}
    D^\ast_\mathrm{cat} = D_\mathrm{v} n_\mathrm{v} f^\ast_\mathrm{v} .
\label{eqn_Dstar}
\end{equation}
Even if $f^\ast_\mathrm{v}$ is reasonably assumed to be of order unity, the division of a $D^\ast_\mathrm{cat}$ value into a $D_\mathrm{v}$ and an $n_\mathrm{v}$ value is only possible if either $D_\mathrm{v}$ or $n_\mathrm{v}$ is independently known.

The problem remains if one only considers the temperature dependence of $D^\ast_\mathrm{cat}$. Writing Arrhenius-type expressions for $D^\ast_\mathrm{cat}$, $D_\mathrm{v}$ and $n_\mathrm{v}$ in Eq.~\eqref{eqn_Dstar}, and assuming $f^\ast_\mathrm{v}$ is temperature independent, one obtains
\begin{equation}
    \Delta H_{D^\ast_\mathrm{cat}} = \Delta H^\ddagger_\mathrm{mig,v} + \Delta H_\mathrm{gen,v}
\end{equation}
where $\Delta H^\ddagger_\mathrm{mig,v}$ is the activation enthalpy of cation-vacancy migration and $\Delta H_\mathrm{gen,v}$ is the generation enthalpy of cation vacancies and indicates how the site fraction of cation vacancies varies with temperature. $\Delta H^\ddagger_\mathrm{mig,v}$ for host cations in \ch{$AB$O3} perovskites is predicted from atomistic simulations to be of the order of several eV \cite{Lewis.1983, Wright.1993, DeSouza.1999, DeSouza.2003, Thomas.2007, Erhart.2007, Walsh.2011, Mizoguchi.2011, Yamamoto.2013, Uberuaga.2013, Lee.2017, Zhang.2018, Parras.2018, Heelweg.2021, Bonkowski.2024, Jackson.2025}. $\Delta H_\mathrm{gen,v}$ can vary from positive values, via zero, to negative values, indicating, respectively, that the site fraction of the relevant defect increases with increasing temperature, is constant as a function of temperature, or decreases with increasing temperature. (In the literature $\Delta H_\mathrm{gen,v}$ is often called the formation enthalpy of cation vacancies, even though $\Delta H_\mathrm{gen,v}$ does not refer to the enthalpy change on removing a cation from the lattice to form a vacancy and even though it can take various values depending on the details of the defect chemistry \cite{DeSouza.2003,Gries.2020}.) Given such large ranges of possible values for $\Delta H^\ddagger_\mathrm{mig,v}$ and $\Delta H_\mathrm{gen,v}$, additional knowledge is required here, too, to divide the measured $\Delta H_{D^\ast_\mathrm{cat}}$ into the two contributions. Thus, answering even the condensed version is also not trivial. In fact, as we will show here, focussing on $\Delta H_{D^\ast_\mathrm{cat}}$ rather than on $D^\ast_\mathrm{cat}$ suffers from a principal weakness.

For cubic \ch{BaTiO3} there are three predicted values of $\Delta H^\ddagger_\mathrm{mig,v_{Ba}}$, one from density-functional-theory (DFT) calculations \cite{Erhart.2007} and two from empirical pair-potential (EPP) calculations \cite{Lewis.1983, Uberuaga.2013}. Normally, the DFT value would be expected to have the highest fidelity, but that is not the case here because the migration barrier was computed by means of molecular static simulations (i.e.~at zero Kelvin) for a higher symmetry structure of \ch{BaTiO3} (cubic) than the ground-state structure (rhombohedral). This is a problem because, in certain defect calculations, relaxation to the ground-state symmetry occurs in addition to relaxation around the point defect(s), and consequently, activation barriers are overestimated \cite{DeSouza.2022}. The value of 6.00 eV reported in Ref.~\citenum{Erhart.2007} must, therefore, be considered an upper limit. (That the activation barriers obtained by DFT calculations\cite{Erhart.2007} for cubic \ch{BaTiO3} are generally overestimated is also evident in the activation enthalpy reported for oxygen-vacancy migration of 0.89~eV being significantly higher than the experimental value of 0.7~eV \cite{Kanert.1994, Kessel.2015, Maier.2016, Cordero.2021}.) The two EPP values of $\Delta H^\ddagger_\mathrm{mig,v_{Ba}}$ may refer to the cubic structure, but they do not clarify the situation because they differ substantially: 3.45 eV \cite{Lewis.1983} vs. 6.68~eV \cite{Uberuaga.2013}. If the respective predictions for the activation barrier of oxygen-vacancy migration (0.62 eV \cite{Lewis.1983} and 1.12 eV \cite{Uberuaga.2013}) are also considered and compared with the experimental value of 0.7~eV,\cite{Kanert.1994, Kessel.2015, Maier.2016, Cordero.2021} it appears that the migration barriers are generally underestimated by \citet{Lewis.1983} and overestimated by \citet{Uberuaga.2013}. Consequently, a reliable value of $\Delta H^\ddagger_\mathrm{mig,v_{Ba}}$ for cubic \ch{BaTiO3} is evidently not available at present.

In this study, we investigated Ba diffusion in cubic \ch{BaTiO3} by means of tracer diffusion experiments and metadynamics simulations. The tracer experiments utilised (i) the stable isotope \ch{^{130}Ba} to probe specifically Ba self-diffusion and (ii) nominally undoped (weakly acceptor-doped) single-crystal samples to ensure high quality and reproducibility as far as sample microstructure and composition are concerned. The diffusion profiles of \ch{^{130}Ba} in the samples were measured by time-of-flight secondary ion mass spectrometry (ToF-SIMS), and their analysis yielded barium tracer diffusion coefficients $D^\ast_\mathrm{Ba}$. Previous studies of Ba tracer diffusion in \ch{BaTiO3} were performed on polycrystalline samples \cite{Garcia.1953, Sazinas.2017}, for which porosity, grain boundaries, and sample roughness may have affected results. Metadynamics (MtD) simulations were employed in order to obtain barium-vacancy diffusion coefficients $D_\mathrm{v_{Ba}}(T)$. As a method of accelerated molecular dynamics, MtD simulations \cite{Laio.2002, Barducci.2011} provide a means for obtaining diffusion coefficients in those cases where high activation barriers result in insufficient ion jumps occurring during standard MD simulations \cite{Koettgen.2018, Ward.2019, Heelweg.2021}. Being based on MD simulations, they also are performed at finite temperatures, and thus, they do not suffer from the problem described above  \cite{DeSouza.2022} that besets zero-Kelvin calculations of migration barriers. In other words, they were carried out at temperatures at which, according to the simulations, cubic \ch{BaTiO3} is the most stable phase. The simulations employed the established set of EPP derived by \citet{Pedone.2006} because such simulations have proven themselves capable of reproducing experimental data. They yield diffusion coefficients as a function of temperature that show excellent quantitative agreement with experiment for oxygen-vacancy diffusion in \ch{BaTiO3} \cite{Kaub.2020}, for oxygen-vacancy diffusion in \ch{SrTiO3} \cite{Waldow.2016}, for oxygen-vacancy diffusion in \ch{CaTiO3} \cite{Robens.2022}, and for strontium-vacancy diffusion in \ch{SrTiO3} \cite{Heelweg.2021}. It is highly likely, therefore, that atomistic simulations based on these EPP will furnish high-fidelity results for cation-vacancy diffusion in \ch{BaTiO3}.

\section{Methods}\label{sec_methods}

\subsection{Sample preparation and diffusion annealing}

Polished, single-crystal samples of \ch{BaTiO3} (CrysTec GmbH, Berlin, Germany) were first annealed in air at the temperature of interest for a time $t_\mathrm{eq}$ to remove polishing damage from the sample surface and to equilibrate the cation sublattices. The samples were then quenched to room temperature. The typical surface roughness, as determined by interference microscopy (Wyco NT1100, Veeco Instruments Inc., Plainview, NY, USA) over an area of $100~\upmu\textrm{m} \times 100~\upmu\textrm{m}$, was (5 to 7) nm. The crystals are considered to be effectively acceptor doped, from the presence of lower-valent impurities and cation vacancies. 

Thin films of \ch{^{130}BaTiO3} were employed as diffusion sources, and they were deposited onto the single crystals by drop coating. A 0.1 M \ch{^{130}BaTiO3} precursor solution was prepared via the acetate route \cite{Hasenkox.1998, Schwartz.2004, Koerfer.2008}, starting with \ch{BaCO3} powder enriched to 32.9\% \ch{^{130}Ba} (Euriso-Top, France) from its natural abundance of 0.11\%, and using titanium tetra-$n$-butoxide. Subsequently, 500 $\upmu$L of the \ch{BaTiO3} precursor solution were mixed together with 0.5 $\upmu$L of surfactant (triethylene glycol monodecyl ether solution) to reduce the surface tension of the precursor solution and improve the wetability of the single-crystal samples. Drop coating was performed in several steps, by dropping 0.95 $\upmu$L of the precursor-surfactant solution onto a single-crystal surface; evaporating the solvent by placing the sample on a hotplate that was held at $T=373$~K; repeating the dropping and evaporation steps; and then, annealing the sample in a pre-heated furnace at $T=973$~K for 20 minutes in order to remove the organic components and form the perovskite structure. This procedure was repeated three further times until a total amount of 7.6 $\upmu$L of the precursor-surfactant solution had been used.

\ch{^{130}Ba} diffusion profiles were produced by annealing the samples at temperatures $1348\leq T/\mathrm{K}\leq 1498$ for diffusion times $220\geq t/\mathrm{h}\geq 49$ (with prior equilibration times $744 \geq t_\mathrm{eq}/\mathrm{h}\geq 240$).

\subsection{SIMS analysis}

Samples were analysed by Time-of-Flight Secondary Ion Mass Spectrometry (ToF-SIMS) on a TOF-SIMS IV machine (IONTOF GmbH, M{\"u}nster, Germany). A beam of 25 keV \ch{Ga+} was scanned over $100~\upmu\textrm{m} \times 100~\upmu\textrm{m}$ of the sample surface to produce secondary ions for ToF analysis. A high-current beam of 2 keV \ch{Cs+} was rastered over $300~\upmu\textrm{m} \times 300~\upmu\textrm{m}$ to sputter etch the sample. Charge compensation was achieved by a beam of low-energy ($<20$ eV) electrons. Negative secondary ions (e.g., \ch{TiO-}, \ch{BaO-} and \ch{O-}) were recorded with a cycle time of 50 $\upmu$s. Crater depths were measured post-analysis with an interference microscope, and the generated craters had depths in the range of (100 to 600) nm. 

\begin{figure}[h]
\centering
\includegraphics{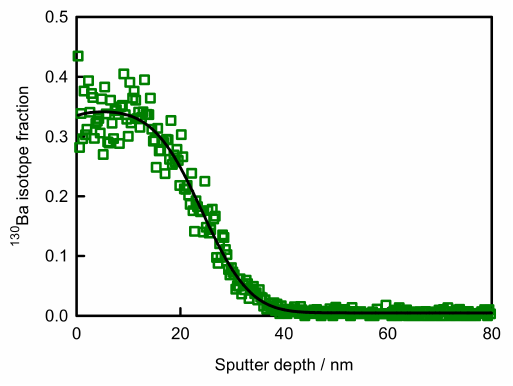}
\caption{ToF-SIMS depth profile of a \ch{^{130}BaTiO3} layer on a single-crystal \ch{BaTiO3} sample prior to diffusion (symbols, experimental data; line, fit of Eq.~\eqref{eqn_diffsoln} for $t = 0$).}\label{fig_zerotime}
\end{figure}

Isotope penetration profiles were analysed with the solution of the diffusion equation for a source of finite thickness (thick film solution) \cite{Crank.1975}, for which the isotope fraction $n^\ast$ as a function of depth $x$ and diffusion time $t$ is given by
\begin{equation}
\label{eqn_diffsoln}
    n^\ast(x,t)=n^\ast_0\left[\mathrm{erf}\left(\frac{x+h}{2\sqrt{D^\ast_\mathrm{Ba}t +\sigma^2}}\right) - \mathrm{erf}\left(\frac{x-h}{2\sqrt{D^\ast_\mathrm{Ba}t +\sigma^2}}\right) \right]
\end{equation}
$n^\ast_0$ is the initial isotope fraction in the deposited layer; $D^\ast_\mathrm{Ba}$ is the tracer diffusion
coefficient of Ba; and $h$ is the film thickness. $\sigma$ describes the apparent broadening of the profile (in addition to the diffusional broadening $\sqrt{D^\ast_\mathrm{Ba}t}$) by a combination of factors: ion beam mixing during sputtering, the roughness of the interface between the film and the single crystal, and any diffusion that may have occurred during the deposition procedure. $\sigma$ and $h$ were determined by analysing the \ch{^{130}Ba} profiles before the diffusion anneals. An example profile is shown in Fig.~\ref{fig_zerotime}, and fitting Eq.~\eqref{eqn_diffsoln} to these data with $t=0$ yields $\sigma = (6.3 \pm 0.5)$~nm and $h = (23.0 \pm 0.5)$~nm. For other profiles $\sigma$ was (5 to 7)~nm and $h$ was (20 to 40)~nm.

\subsection{Metadynamics (MtD) simulations}

If it were possible to investigate Ba diffusion in perovskite \ch{BaTiO3} with standard MD simulations, some number of barium vacancies would be introduced into a simulation cell, and the Ba tracer diffusion coefficient would be extracted from the evolution of the mean square displacement of all barium ions as a function of time \cite{Bonkowski.2025}. In a MtD simulation \cite{Koettgen.2018, Ward.2019, Heelweg.2021} a single barium ion is forced through the stepwise addition of Gaussian bias potentials, acting on selected degrees of freedom of the system (termed collective variables, CVs), to overcome the free energy barrier and jump to an adjacent vacancy. 

Following \citet{Chandler.1978}, the rate at which the free energy barrier is crossed, i.e.~the successful jump rate $\Gamma$, can be expressed within classical transition-state theory as \cite{MSbook.2002}
\begin{equation}
\label{eqn_Gammajump}
    \Gamma = \frac{\langle |\dot{\xi} | \rangle_{\xi(0)=\xi^{\ast}}}{2} P(\xi^{i}) \exp\left(- \frac{\Delta G_{\mathrm{mig,v}}^{\ddagger}}{k_{\mathrm{B}} T}\right).
\end{equation}
$\langle |\dot{\xi} | \rangle_{\xi(0)=\xi^{\ast}}$ is the ensemble average of the velocity observable $|\dot{\xi}|$ for the case that the system is initially ($t=0$) held at $\xi^\ast$; $P(\xi^{i})$ is the probability density of finding the system in the initial configuration, $\xi^i$; and $\Delta G_{\mathrm{mig,v}}^{\ddagger}$ is the Gibbs energy difference between the system at the barrier top $\xi^\ast$ and the initial configuration, i.e., the activation Gibbs energy of migration. Essentially, Eq.~\eqref{eqn_Gammajump} can be understood as the product of the mean velocity of the system to cross the barrier without falling back to the initial potential well (as given by the velocity average divided by 2 to only include successful crossings) and the actual probability of finding the system at that barrier maximum, which is given by the product of $P(\xi^{i})$ and the exponential term.

From random-walk theory, the diffusion coefficient of barium vacancies is then obtained according to 
\begin{equation}
    D_{\mathrm{v_{Ba}}} = \frac{Z}{6} d_{\mathrm{v}}^2 \Gamma ,
\end{equation}
where $Z$ is the number of nearest jump neighbours and $d_{\mathrm{v}}$ is the jump distance.

$\Delta G_{\mathrm{mig,v}}^{\ddagger}$ was computed by means of MtD simulations. $P(\xi^i)$ was obtained from straightforward MD simulations by sampling the probability density around the initial position. And $\langle |\dot{\xi} | \rangle_{\xi(0)=\xi^{\ast}}$ was calculated by constrained MD simulations: the migrating ion was first confined with an harmonic potential to the top of the barrier ($\xi^{\ast}$), the confinement was removed abruptly, and the velocity of the migrating ion leaving $\xi^{\ast}$ for one of the two adjacent minima was then determined. Averaging over a large number of simulation runs (in this case, 1000) yielded $\langle |\dot{\xi} | \rangle_{\xi(0)=\xi^{\ast}}$.

All simulations were performed with the \textsc{Lammps} code \cite{Thompson.2022}. An $8a \times 8a \times 8a$ supercell of \ch{BaTiO3} was employed, where $a$ is the lattice parameter of cubic \ch{BaTiO3}. The classical equations of motion were solved with the velocity Verlet algorithm, with a time step of $\Delta t =1$~fs. The isothermal--isobaric ($NpT$) ensemble was used, and temperature $T$ and pressure $p=0$ were controlled by means of a Nos\'e--Hoover thermostat and barostat with relaxation times of 100 and 1000 time steps. Periodic boundary conditions were applied. Coulomb interactions were computed with the particle-particle particle-mesh method, with an accuracy of ${10^{-6}}$. For the MtD simulations, the \textsc{Colvars} package \cite{Fiorin.2013} as implemented in \textsc{Lammps} was used. Spatial coordinates in the Cartesian coordinate system were employed as the CVs, and Gaussian bias potentials of height $0.005$~eV and width $2$~\AA\ were added to the system every 100 time steps.

Regardless of the outstanding agreement with experimental data for defect diffusion in titanate perovskites, as noted in the Introduction, it is worth asking the question, because the study relies strongly on the fidelity of the predicted defect diffusion coefficients, as to why the EPP derived by \citet{Pedone.2006} were employed rather than DFT calculations or even a universal machine-learning (ML) potential. If one had decided to perform DFT calculations, one would have to choose carefully the DFT functional, since migration barriers can vary substantially depending on the functional used \cite{Zhang.2016, Devi.2022}. Furthermore, one could not use zero-Kelvin calculations to compare results from different functionals --- they will not provide reliable results\cite{DeSouza.2022} --- but one would have to carry out MtD simulations as a function of temperature. This would be an enormous computational expense. As for an off-the-shelf ML potential, it is highly unlikely to reproduce experimental diffusion data for a titanate perovskite \cite{Perkampus.2025} without careful tuning, and as the tuning data is generated by DFT calculations, here, too, the choice of DFT functional is critical.

\section{Results and Discussion}\label{sec_results}

\subsection{Tracer diffusion experiments}

\begin{figure}[h]
\centering
\includegraphics{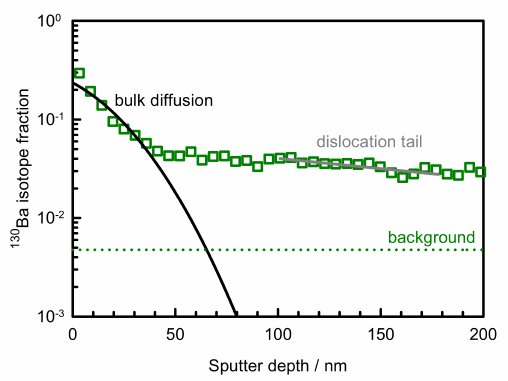}
\caption{\ch{^{130}Ba} tracer diffusion profile obtained by ToF-SIMS depth profiling after a diffusion anneal in air at $T = 1448$~K for $t = 95$~h (symbols, experimental data; black line, fit of Eq.~\eqref{eqn_diffsoln} for bulk diffusion; grey line, fit to $\ln n^\ast \propto x^1$ for faster dislocation diffusion).}\label{fig_diffusionprofile}
\end{figure}

An exemplary $^{130}$Ba penetration profile obtained by ToF-SIMS depth profiling of a single-crystal \ch{BaTiO3} sample after diffusion annealing is shown in Fig.~\ref{fig_diffusionprofile}. Two features are observable above the background detection level. The feature closer to the surface is ascribed to bulk diffusion, and it can be described by the appropriate solution to the diffusion equation [see Eq.~\eqref{eqn_diffsoln}] to yield $D^\ast_\mathrm{Ba}$. The feature at larger depths is ascribed to faster diffusion along dislocations, because the single-crystal samples had neither
grain boundaries nor connected porosity, and because the profile feature has the appropriate mathematical form, $\ln n^\ast \propto x^1$ \cite{LeClaire.1981}. The observation of faster cation diffusion along dislocations in an acceptor-doped perovskite accords well with results obtained for Sr and Zr impurity diffusion in \ch{BaTiO3} \cite{Koerfer.2008}, Sr tracer diffusion in \ch{SrTiO3} \cite{Rhodes.1966} and for Dy and Sc impurity diffusion in \ch{LaInO3} \cite{Liedtke.2025}. Dislocation diffusion coefficients of Ba will be presented in a future paper, and the possible role played by space-charge tubes in giving rise to faster dislocation diffusion of cations will be discussed \cite{DeSouza.2021, Parras.2020}.

\begin{figure}[h]
\centering
\includegraphics{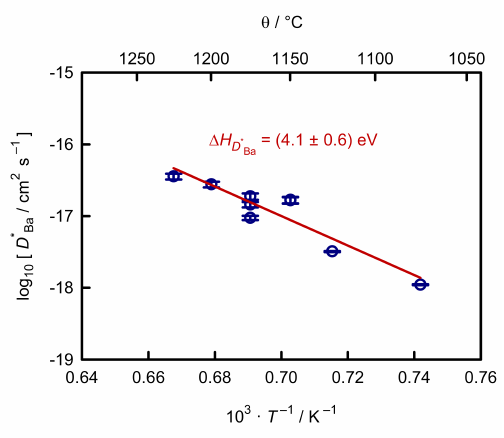}
\caption{\ch{^{130}Ba} tracer diffusion coefficients in single-crystal \ch{BaTiO3} obtained experimentally by ToF-SIMS depth profiling as a function of inverse temperature. .}\label{fig_Dstar}
\end{figure}

Tracer diffusion coefficients of Ba obtained for the bulk as a function of temperature are shown in Fig.~\ref{fig_Dstar}, and their analysis yields an activation enthalpy of Ba tracer diffusion of $\Delta H_{D_\mathrm{Ba}}^\ast = (4.1\pm0.6)$~eV. This value is consistent with the majority of literature data [(4 to 6)~eV] summarised in Fig.~\ref{fig_summaryliterature} for cation diffusion in \ch{BaTiO3}.

\subsection{Metadynamics simulations}

$D_\mathrm{v_{Ba}}(T)$ were computed for barium-vacancy migration occurring: (i) by an isolated barium vacancy, (ii) as part of a defect associate with an oxygen vacancy, and (iii) as part of a defect associate with a titanium vacancy. The two latter possibilities were investigated because in \ch{$AB$O3} perovskites the $\mathrm{(v_\textit{A}v_{O})}$ associate \cite{Walsh.2011, Yamamoto.2013, Heelweg.2021} and the $\mathrm{(v_\textit{A}v_\textit{B})}$ associate \cite{DeSouza.1999, DeSouza.2003, Mizoguchi.2011, Yamamoto.2013, Lee.2017, Bonkowski.2024} have been predicted to have lower activation enthalpies of $A$ migration than that of the isolated vacancy. $\mathrm{(v_\textit{A}v_\textit{B})}$ associates, in particular, provide an explanation for $\Delta H_{D_A} \approx \Delta H_{D_B}$ (see, e.g., Fig.~\ref{fig_summaryliterature}) with ${D_A} \approx {D_B}$ (see, e.g., Refs.~\citenum{Koerfer.2008, Yoo.2008, Lee.2008} for \ch{BaTiO3}). $\mathrm{(v_{Ba}v_O)}$ associates are expected to be bound on account of the opposite relative charges of the constituent defects, and DFT calculations \cite{Erhart.2007} confirm this expectation. The association energy of the $\mathrm{(v_{Ba}v_{Ti})}$ pair has not been examined, but against expectations it may be only slightly unfavourable at worst (cf.~results for \ch{LaGaO3} \cite{DeSouza.2003} and \ch{LaMnO3} \cite{Lee.2017}); it may also become favourable through the addition of one or more oxygen vacancies, forming for instance the defect cluster $\mathrm{(v_{Ba}v_{Ti}v_O)}$ \cite{Schulz.2003, Belova.2007}, but such complicated variants were not examined in the simulations. Given the importance of Coulomb interactions between the migrating $A$-site cation and the surrounding $B$-site cations in determining the activation enthalpy of $A$ migration \cite{DeSouza.2003}, it seems unlikely that additional $\mathrm{v_{O}}$ will change the results substantially.

\begin{figure}[h]
\centering
\includegraphics{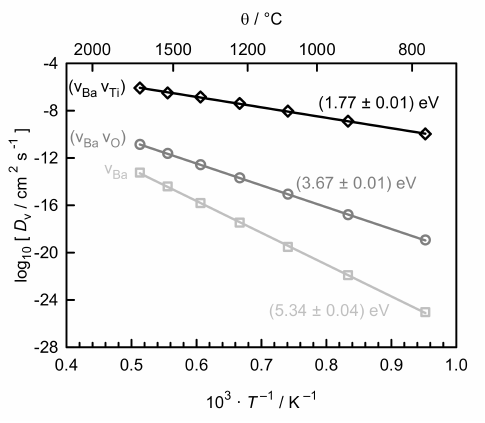}
\caption{Diffusion coefficients of barium vacancies in cubic \ch{BaTiO3} as a function of inverse temperature obtained by metadynamics simulations for barium-vacancy migration by isolated $\mathrm{v_{Ba}}$, as part of $\mathrm{(v_{Ba}v_{O})}$, or as part of $\mathrm{(v_{Ba}v_{Ti})}$. }\label{fig_DvBa}
\end{figure}

The results of the simulations are plotted in  Fig.~\ref{fig_DvBa}. In the order [$\mathrm{v_{Ba}}$, $\mathrm{(v_{Ba}v_{O})}$, $\mathrm{(v_{Ba}v_{Ti})}$], isothermal values of $D_\mathrm{v_{Ba}}$ are seen to increase, and in this order, the corresponding activation enthalpies of migration decrease [5.34~eV, 3.67~eV, 1.77~eV].

Since we only examined in the metadynamics simulations the jump of the barium cation explicitly, in calculating a barium-vacancy diffusion coefficient, we implicitly assumed that the jump rate of the other defect in the associate is not vastly slower. While this is a reasonable assumption for the $\mathrm{(v_{Ba}v_O)}$ associate, it is less so for the $\mathrm{(v_{Ba}v_{Ti})}$ associate. For this reason, additional metadynamic simulations were carried out for Ti migration as part of the $\mathrm{(v_{Ba}v_{Ti})}$ associate. The predicted diffusion coefficients for Ti migration (not shown) are almost identical to those of Ba migration. In other words, both cations have similar jump rates as part of the $\mathrm{(v_{Ba}v_{Ti})}$ associate.

In order to determine which of these mechanisms is responsible for the experimental diffusion data of Fig.~\ref{fig_Dstar}, we focussed on the absolute magnitude of the diffusivities, and not their activation enthalpies. To be specific, we took from Fig.~\ref{fig_Dstar} that the \ch{^{130}Ba} tracer diffusivity at $T = 1430$~K is $D^\ast_\mathrm{Ba}\approx10^{-17}$~cm$^2$~s$^{-1}$. Extracting the $D_\mathrm{v_{Ba}}$ values for $T = 1430$~K from Fig.~\ref{fig_DvBa}, we calculated with Eq.~\eqref{eqn_Dstar} the site fraction of defects that is required to reproduce the experimental value, assuming $f^\ast_\mathrm{def} =1$ for all three cases. ($n_\mathrm{def}$ is relative to the volumentric density of Ba cation sites in all three cases.) For the three mechanisms [$\mathrm{v_{Ba}}$, $\mathrm{(v_{Ba}v_{O})}$, $\mathrm{(v_{Ba}v_{Ti})}$], we thus obtain site fractions of $n_\mathrm{def} = 10^{1.37}, 10^{-2.72}, 10^{-9.30}$. Since $n_\mathrm{def} > 1$ is obtained for the first case---a physical impossibility: there cannot be more vacancies than sites than can be vacant---, we can unequivocally rule out the migration of isolated barium vacancies as the predominant diffusion mechanism for Ba in these samples. The other two mechanisms are both possible, since $n_\mathrm{def} < 1$ is found. If the second mechanism of $\mathrm{(v_{Ba}v_{O})}$ migration is dominant, then the third mechanism cannot be, and this requires the site fraction of $\mathrm{(v_{Ba}v_{Ti})}$ at $T = 1430$~K to be below $10^{-9.30}$; otherwise, the substantially higher diffusivity of $\mathrm{(v_{Ba}v_{Ti})}$ associates would provide the main contribution to the measured cation diffusion coefficient. To emphasise this point, we consider the hypothetical case that at $T = 1430$~K the site fractions of [$\mathrm{v_{Ba}}$, $\mathrm{(v_{Ba}v_{O})}$, $\mathrm{(v_{Ba}v_{Ti})}$] are [$n_\mathrm{def} = 10^{-3}, 10^{-6}, 10^{-10}$]: the dominant contribution to the Ba diffusivity would come from the $\mathrm{(v_{Ba}v_{Ti})}$ associates on account of their extremely high defect diffusivites. It is important to note that, although $f_\mathrm{def}^\ast$ is only known for the $\mathrm{v_{Ba}}$ mechanism (the Ba cations form a simple cubic sublattice, so that $f_\mathrm{v}^\ast=0.65$), including $f_\mathrm{def}^\ast$ values in the analysis would not alter the main conclusions. Given that $f_\mathrm{def}^\ast < 1$ is expected in all three cases, the inclusion of $f_\mathrm{def}^\ast$ values will increase the $n_\mathrm{def}$ values at $T = 1430$~K, each to a different degree, but the order will remain unchanged. We cannot conclude, however, that diffusion definitely occurs by one of these associate mechanisms, since we have not examined all possible migration mechanisms and ruled out all possibilities except for these two. Nevertheless, in the absence of reasonable alternatives, we propose that the two associate mechanisms (including the related cluster mechanism \cite{Schulz.2003, Belova.2007}) are the leading contenders.

NB: If one were to use the standard approach in the literature, that is, the comparison of activation enthalpies, one would in all probability reach a different conclusion. As one would only require an activation enthalpy from simulations, MtD simulations would not be necessary --- a simple molecular static calculation would suffice. This would yield, for example, with the Pedone EPP\cite{Pedone.2006} that we used here, $\Delta H_\mathrm{mig,v}^\ddagger = 5.3$~eV  for the migration of isolated $\mathrm{v_{Ba}}$. Consequently, with $\Delta H_{D_\mathrm{Ba}}^\ast = (4.1\pm0.6)$~eV, one would conclude that, within experimental error of two standard deviations ($\pm 1.2$~eV), there is agreement between the experimental value and the computational prediction (with $\Delta H_\mathrm{gen,v}$ being essentially zero). There would be no need, therefore, to examine alternative mechanisms (as we have done here). Hence, by focussing on activation enthalpies one would reach an incorrect conclusion, one that is inconsistent with the absolute magnitude of the diffusivities.

\subsection{Comparison with literature}

\begin{figure}[h]
\centering
\includegraphics{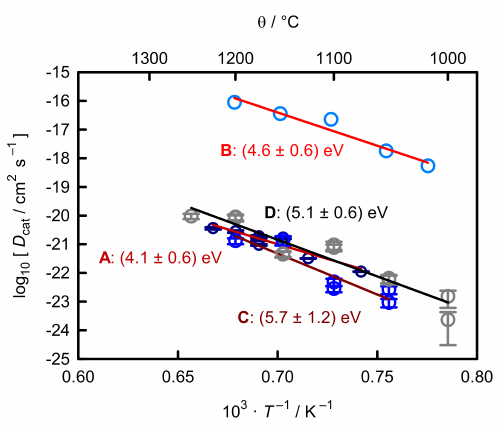}
\caption{Comparison of selected cation diffusion coefficients for cubic \ch{BaTiO3} obtained experimentally as a function of inverse temperature: A, \ch{^{130}Ba} tracer diffusion (this study); B, \ch{^{134}Ba} tracer diffusion \cite{Sazinas.2017}; C: Sr impurity diffusion \cite{Koerfer.2008}; D: Zr impurity diffusion \cite{Koerfer.2008}.}\label{fig_Dcatcomp}
\end{figure}

It is instructive to compare our tracer diffusion coefficients with literature values for Ba tracer diffusion \cite{Sazinas.2017} and for Sr impurity ($A$-site) diffusion and Zr impurity ($B$-site) diffusion \cite{Koerfer.2008} in cubic \ch{BaTiO3}. We focus first on diffusion of $A$-site cations. In Fig.~\ref{fig_Dcatcomp} we see that at $T = 1430$~K, for instance, in comparison with our value of $D^\ast_\mathrm{Ba}\approx10^{-17}$~cm$^2$~s$^{-1}$, the $D^\ast_\mathrm{Ba}$ value obtained by \citet{Sazinas.2017} is higher by a factor of $\approx 10^{4.6}$, whereas the $D_\mathrm{Sr}$ value obtained by \citet{Koerfer.2008} is slightly lower (by a factor of $\approx 2$). It is noted that \citet{Koerfer.2008} also employed single-crystal samples (albeit from a different supplier), while \citet{Sazinas.2017} used polycrystalline samples. However, we do not believe that the observed differences arise from the use of polycrystalline samples versus single crystals: rather we expect the differences to arise from the samples' differing defect populations: each set of samples is likely to have contained differing amounts of $\mathrm{v_{Ba}}$, $\mathrm{(v_{Ba}v_{O})}$ and $\mathrm{(v_{Ba}v_{Ti})}$ defects.

Performing the same comparative analysis of $D_\mathrm{cat}$ and $D_\mathrm{v_{Ba}}$ as above, we ascertain that, as $n_\mathrm{def}>1$ is obtained, in neither case \cite{Koerfer.2008, Sazinas.2017} can the migration of isolated $\mathrm{v_{Ba}}$ be responsible for diffusion of $A$-site cations. Furthermore, we conclude that, of the two other possibilities, for the $D_\mathrm{Sr}$ data of \citet{Koerfer.2008} both migration mechanisms involving defect associates are possible, but for the $D^\ast_\mathrm{Ba}$ data of \citet{Sazinas.2017} only the $\mathrm{(v_{Ba}v_{Ti})}$ mechanism is physically possible.

Now we include the $D_\mathrm{Zr}$ data in our comparison. Since \citet{Koerfer.2008} found ${D_\mathrm{Sr}} \approx {D_\mathrm{Zr}}$, with $\Delta H_{D_\mathrm{Sr}} \approx \Delta H_{D_\mathrm{Zr}}$, this strongly suggests that the $D_\mathrm{Sr}$ data (and the $D_\mathrm{Zr}$ data) arise from the migration of $\mathrm{(v_{Ba}v_{Ti})}$ associates [or defect clusters, such as $\mathrm{(v_{Ba}v_{Ti}v_O)}$ that contain the $\mathrm{(v_{Ba}v_{Ti})}$ associate]. The alternative is that $\mathrm{(v_{Ba}v_{O})}$ migration yields values for $D_\mathrm{Sr}$ and $\Delta H_{D_\mathrm{Sr}}$ that are very similar to those produced by a completely different mechanism, such as $\mathrm{v_{Ti}}$ or $\mathrm{(v_{Ti}v_{O})}$ migration, for $D_\mathrm{Zr}$ and $\Delta H_{D_\mathrm{Zr}}$. Although it cannot be ruled out entirely, we consider this alternative to be highly unlikely.

If all four sets of data are thus considered together, there are, in effect, two scenarios. The first scenario has all four datasets (A, B, C and D) sharing the same mechanism of migration: cation diffusion occurs by means of $\mathrm{(v_{Ba}v_{Ti})}$ associates. Each dataset is consistent, within error, with $\Delta H_D\approx 4.5$~eV, and the differences between the datasets arise from differing site fractions of $\mathrm{(v_{Ba}v_{Ti})}$ associates (at $T = 1430$~K, the variation is, then, $\approx10^5$). These differences are due possibly to differing levels of background acceptor impurities, differing Ba/Ti ratios, or differing degrees of thermodynamic equilibration. Assuming reasonably that $f_\mathrm{def}^\ast$ does not exhibit a strong dependence on temperature (cf.~Ref.~\citenum{divinski.2000}), one obtains with $\Delta H^\ddagger_\mathrm{mig,v}= 1.8$~eV that the site fraction of these associates varies strongly with temperature, $\Delta H_\mathrm{gen,v} = \Delta H_{D_\mathrm{Ba}^\ast} - \Delta H_\mathrm{mig,v}^\ddagger= 2.7$~eV. The second scenario, consequently, is that the migration mechanisms are not all the same for all three: for dataset A, diffusion occurs by a $\mathrm{(v_{Ba}v_{O})}$ mechanism, while for datasets B, C and D, the $\mathrm{(v_{Ba}v_{Ti})}$ mechanism prevails. We deem it unlikely but not impossible that both $D_\mathrm{Ba}^\ast$ and $\Delta H_{D^\ast_\mathrm{Ba}}$ (dataset A) produced by one mechanism are comparable to $D_\mathrm{cat}$ and $\Delta H_{D_\mathrm{cat}}$ (datasets C and D) from a different mechanism. Further conclusions are only possible with further data. In particular, barium, titanium and oxygen tracer diffusion experiments as a function of temperature and Ba/Ti ratio would be helpful. 

For a range of Ba/Ti ratios around 1, the main mechanism of cation diffusion in \ch{BaTiO3} will involve $\mathrm{(v_{Ba}v_{Ti})}$ associates [or defect clusters\cite{Schulz.2003, Belova.2007} containing the $\mathrm{(v_{Ba}v_{Ti})}$ associate]. Heading towards extreme Ti-rich compositions, the dominant mechanism will at some point become $\mathrm{(v_{Ba}v_{O})}$ or $\mathrm{v_{Ba}}$ migration, as the concentration of $\mathrm{(v_{Ba}v_{Ti})}$ associates will be so many orders of magnitude lower as to provide no significant contribution. For Ba-rich compositions, cation diffusion will be very slow, as the concentration of $\mathrm{(v_{Ba}v_{Ti})}$ associates will be very low, and the alternative mechanisms [involving $\mathrm{v_{Ti}}$ or $\mathrm{(v_{Ti}v_{O})}$ migration] probably provide no significant contributions.

\section{Conclusions}

We studied the diffusion of Ba in cubic \ch{BaTiO3} using tracer diffusion experiments and metadynamics simulations.The main conclusions to emerge from our study are as follows:
\begin{itemize}
    \item Analysis of \ch{$^{130}$Ba} penetration profiles into single-crystal samples of nominally un-doped (effectively acceptor-doped) \ch{BaTiO3} revealed slow bulk diffusion of Ba cations (the subject of this study) and faster diffusion along dislocations (the subject of a coming study). Tracer diffusion coefficients of Ba obtained for temperatures $1348 \leq T/\mathrm{K} \leq 1498$ were $10^{-18.0} \leq D_\mathrm{Ba}^\ast/\mathrm{cm^2\; s^{-1}} \leq 10^{-16.5}$.
    
    \item Diffusion coefficients of barium vacancies, $D_\mathrm{v_{Ba}}$, were obtained as a function of temperature by metadynamics simulations for three mechanisms --- migration by means of $\mathrm{v_{Ba}}$, $\mathrm{(v_{Ba}v_{O})}$, or $\mathrm{(v_{Ba}v_{Ti})}$. Isothermal values of $D_\mathrm{v_{Ba}}$ are found to increase in the order $\mathrm{v_{Ba}}$, $\mathrm{(v_{Ba}v_{O})}$, $\mathrm{(v_{Ba}v_{Ti})}$. The data refer unambiguously to the cubic phase of \ch{BaTiO3}.

    \item The extremely high defect diffusivities predicted for $\mathrm{(v_{Ba}v_{Ti})}$ migration relative to those for $\mathrm{v_{Ba}}$ or $\mathrm{(v_{Ba}v_{O})}$ migration translate into very few $\mathrm{(v_{Ba}v_{Ti})}$ associates being necessary for this mechanism to provide the dominant contribution to $D_\mathrm{Ba}^\ast$.
    
    \item The combined analysis of $D_\mathrm{Ba}^\ast$ and $D_\mathrm{v_{Ba}}$ obtained in this study, together with literature data \cite{Koerfer.2008,Sazinas.2017} for $D_\mathrm{cat}$, indicates that Ba tracer diffusion, as probed in this study, cannot occur by the migration of $\mathrm{v_{Ba}}$, may occur by the migration of $\mathrm{(v_{Ba}v_{O})}$ associates, and most likely occurs by the migration of $\mathrm{(v_{Ba}v_{Ti})}$ associates [or some larger defect cluster\cite{Schulz.2003, Belova.2007} containing $\mathrm{(v_{Ba}v_{Ti})}$].
    
    \item The Ba/Ti ratio in \ch{BaTiO3} may strongly affect the effective diffusivities of barium vacancies (and titanium vacancies), and not only the concentrations of these defects.

    \item Attention is drawn to the danger of the standard approach to analysing diffusion data, that is, the inspection of activation enthalpies ($\Delta H_{D^\ast_\mathrm{Ba}}$, $\Delta H_\mathrm{mig,v}^\ddagger$) and thus to the benefit of focussing on diffusion coefficients ($D_\mathrm{Ba}^\ast$, $D_\mathrm{v_{Ba}}$).

\end{itemize}

\begin{acknowledgement}

This study has received funding from the Deutsche Forschungsgemeinschaft (DFG, German Research Foundation) within projects 5248730 and 463184206 (SFB 1548, FLAIR: Fermi Level Engineering Applied to Oxide Electroceramics). Simulations were performed with computing resources granted by RWTH Aachen University under projects rwth0753 and rwth1744.

\end{acknowledgement}

\bibliography{BaTiO3}

\end{document}